\documentclass[a4paper,twocolumn,11pt,allowtoday]{quantumarticle}

\pdfoutput=1
\usepackage[utf8]{inputenc}
\usepackage[english]{babel}
\usepackage[T1]{fontenc}
\usepackage{amsmath}
\usepackage{amssymb}
\usepackage{hyperref}
\usepackage{tikz}
\usepackage{lipsum}
\usepackage[numbers]{natbib}
\usepackage[dvipsnames]{xcolor}
\usepackage{array}
\usepackage{graphicx}
\usepackage{physics}

\begin{document}
\title{Information-efficient decoding of surface codes}

\author{Long DH My}
\email{long@entropicalabs.com}
\affiliation{Entropica Labs, Singapore}

\author{Shao-Hen Chiew}
\thanks{Current address: Institute of Physics, École Polytechnique Fédérale de Lausanne (EPFL), CH-1015 Lausanne, Switzerland; Center for Quantum Science and Engineering, École Polytechnique Fédérale de Lausanne (EPFL), CH-1015 Lausanne, Switzerland.}
\affiliation{Entropica Labs, Singapore}

\author{Jing Hao Chai}
\affiliation{Entropica Labs, Singapore}

\author{Hui Khoon Ng}
\email{huikhoon.ng@nus.edu.sg}
\affiliation{Entropica Labs, Singapore}
\affiliation{Department of Physics, National University of Singapore, Singapore}
\affiliation{Centre for Quantum Technologies, National University of Singapore, Singapore}

\maketitle

\begin{abstract}
Surface codes are a popular error-correction route to fault-tolerant quantum computation. The so-called exponential backlog problem that can arise when one has to do logical $T$-gates within the surface code demands real-time decoding of the syndrome information to diagnose the appropriate Pauli frame in which to do the gate. This in turn puts a minimum requirement on the communication rate between the quantum processing unit, where the syndrome information is collected, and the classical processor, where the decoding algorithm is run. This minimum communication rate can be difficult to achieve while preserving the quality of the quantum processor. Here, we present two decoders that make use of a reduced syndrome information volume, relying on a number of syndrome bits that scale only as the width---and not the usual area---of the surface-code patch. This eases the communication requirements necessary for real-time decoding.
\end{abstract}

\section{Introduction}

Surface codes \cite{kitaev97,kitaev03,dennis02} are the most well studied among error-correcting codes for quantum computing. That nearest-neighbor interactions suffice, coupled with the not-too-stringent noise threshold, have made the planar variant \cite{bravyi98,fowler12} the code of choice for many hardware architectures. The past three years have seen experimental demonstrations of beyond-breakeven error correction using surface codes \cite{google2023suppressing, eickbusch2025demonstration, google2025quantum, bluvstein2024logical, bluvstein2025architectural}. The success of surface codes also motivated the current intense search for good quantum low-density parity-check (q-LDPC) codes with similar locality and threshold properties but with better coding rates \cite{breuckmann2021,bravyi2024high, xu2024constant, pattison2025hierarchical, berthusen2025toward}.

Error correction with surface codes, as is true with stabilizer codes in general, demands repeated measurements of the stabilizer generators that define the code. These measurements generate syndrome bits across the entire code lattice, and these bits are processed by a classical decoder for error diagnosis. Many decoding algorithms exist, including the traditional minimum-weight perfect-matching (MWPM) via the Blossom algorithm \cite{edmonds1965} (implemented in the popular PyMatching package \cite{higgott22}), the union-find approach \cite{delfosse2021}, the Sparse-Blossom variant of MWPM \cite{higgott25}, the maximum-likelihood decoder \cite{bravyi2014}, as well as machine-learning decoding protocols (e.g., Refs.~\cite{torlai2017,krastanov2017}); see also Ref.~\cite{iOlius2024} for a review on surface-code decoders. Each of these decoding algorithms takes in the measured syndrome bits and produces a best guess for the errors that have occurred to generate those syndromes. 

For use within a quantum computation, the decoders must be able to keep up with the syndrome generation, to not suffer from the so-called backlog problem \cite{terhal2015quantum} that can lead to exponential slowdown of the computation, negating any quantum speed advantage one may hope to achieve. Syndromes have to be processed as they are generated \cite{terhal2015quantum, skoric2023parallel, tan2023scalable}, to deduce the correct Pauli frame needed at the moment of application of a logical $T$ gate (via the magic-state approach \cite{bravyi2005}) during the computation. This puts a requirement on the classical communication rate between the quantum processing unit and the classical computer that performs the decoding. 

Past works \cite{das_afs_2022, zhang_classical_2023,camps_evaluation_2024} have analyzed the communication bandwidth requirements. In every syndrome measurement cycle, for each logical qubit encoded with distance-$d$ rotated planar surface code, $O(d^2)$ bits of syndrome information are generated---the number of syndrome bits scales as the area of the $d\times d$ lattice. The transmission of the $d^2$ bits of information per logical qubit has to be completed within the syndrome measurement cycle time to keep up with the syndrome generation rate. This translates into gigabits or even terabits per second of classical data transmission for typical noise and reasonably sized quantum algorithms \cite{das_afs_2022}; for comparison, a typical server stack can handle perhaps terabits per second of communication. Handling such large data transmission rates requires multiple high-bandwidth classical communication channels, but too many such connections between the (typically) cryogenic quantum processing unit (QPU) where the syndromes are generated and the hot classical processing unit (CPU) outside will ultimately degrade the quality of the QPU.

Ref.~\cite{das_afs_2022} proposed data compression protocols to reduce this requisite data transmission rate, noting that nontrivial syndromes arise at a rate that scales with the error probability. Yet, any such data compression, to actually reduce the data transmission rate, has to be done within the QPU, hence limiting the protocols to only those simple enough for the on-chip CPU (perhaps an FPGA with only basic Boolean operations) to handle. Furthermore, one cannot guarantee efficient compression in all cases since the syndrome data can, in principle, be completely random.

We thus ask the question, before we attempt such data compression, if all the syndrome data are necessary for decoding in the first place. Can we have more information-efficient decoders instead? Here, we answer in the affirmative for planar surface codes, and introduce two information-efficient decoders: the row-column decoder and the boundary decoder. The row-column decoder aggregates the syndrome information in a row or column---simple enough for an on-chip FPGA to handle---and only those aggregates are transmitted to and processed by the classical decoding algorithm (e.g., MWPM). The boundary decoder relies on the dynamic syndrome measurement circuits of Ref.~\cite{mcewen23} to ``pump" errors to the boundaries of the surface-code lattice such that only the syndrome bits at the boundaries have to be extracted to the decoding algorithm for error diagnosis. In both cases, the number of transmitted syndrome bits scales with $d$, the width of the surface-code lattice, rather than with its area $d^2$.

As we show below, both information-efficient decoders preserve the code distance. They do, however, give worse logical error probabilities compared to the standard---or ``areal"---decoders that make use of the syndrome data across the whole lattice; after all, the information-efficient decoders rely only on a genuine subset of available syndrome data. Surface codes are well known to correct many more error patterns than promised by the code distance, by relying on the fact that errors that are well distributed across the entire surface can be dealt with correctly, even if their number exceeds the code capacity. Our row-column and boundary decoders discard such areal location information and hence cannot retain that feature of surface codes. This does not, however, preclude the possibility of inventing information-efficient decoders with improved logical error probabilities. The dynamic syndrome measurement circuit that underlies the boundary decoder method, in particular, opens the door to schemes that can potentially pump errors across the lattice in a manner so that errors self cancel without requiring decoding. 

Below, we begin with a quick review of the necessary surface-code concepts. Next, we describe the two information decoders, first the row-column decoder, followed by the boundary decoder. We then investigate the efficacy of the decoders in numerical memory experiments, comparing the resulting logical error probability with that from standard decoders.

\section{Planar surface code basics}

\begin{figure}
\includegraphics[trim=8mm 173mm 28mm 50mm, clip, width=1\columnwidth]{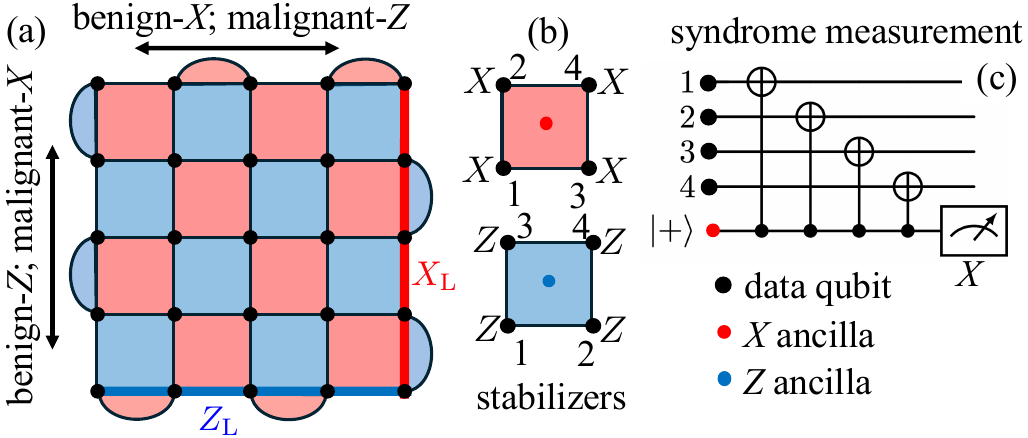}
\caption{\label{fig:SurfCodePatch} Standard rotated planar surface code. (a) A surface-code patch carrying a single logical qubit, with $X_\mathrm{L}$ (vertical red line) and $Z_\mathrm{L}$ (horizontal blue line) as the logical $X$ and $Z$ operators (b) The code stabilizers (the generators), $XXXX$ (red) or $ZZZZ$ (blue) on the data qubits at the corners of the square, tiled across the surface-code patch; boundary stabilizers [semicircular regions in (a)] are supported only on pairs of data qubits. (c) Standard syndrome measurement circuit for the $X$ stabilizer using a single ancillary qubit; the corresponding circuit for the $Z$ stabilizer is identical, except for an $X\leftrightarrow Z$ swap---the $Z$ ancilla starts in $\ket{0}$, the four CNOTs have their targets and controls switched, and the ancilla is measured in the $Z$ basis.}
\end{figure}

A rotated planar surface code patch comprises a checkerboard pattern of $XXXX$ and $ZZZZ$ stabilizers ($X$ and $Z$ stabilizers for short), each supported on the four physical---``data''---qubits at the corner of each square, with those at the boundaries having smaller support; see Fig.~\ref{fig:SurfCodePatch}. A patch, built from $d^2$ ($d$ an odd positive integer) data qubits, carries a single logical qubit in a chosen common (two-dimensional) eigenspace of all the $X$ and $Z$ stabilizers, and can correct arbitrary errors on up to $t\equiv \lfloor \frac{1}{2}(d-1)\rfloor$ data qubits. The logical qubit is manipulated using logical operators that commute with all the code stabilizers. A standard choice is to have the logical $X$ operator extends across the patch in one direction [vertically in Fig.~\ref{fig:SurfCodePatch}(a)] while the logical $Z$ operator extends in an orthogonal direction (horizontally in the figure).

To correct errors, the $X$ and $Z$ stabilizers are measured across the entire surface-code patch, with each stabilizer measurement---a syndrome measurement---employing a single ancillary qubit coupled to surrounding data qubits via CNOT gates; see the circuits that carry out syndrome extraction (SE) in Fig.~\ref{fig:SurfCodePatch}(c). A nontrivial syndrome---a defect---arises whenever there is a flip in the eigenvalue of a stabilizer, and the defect location is noted. The $X(Z)$ stabilizers detect $Z(X)$ errors in the data qubits. Assuming error-free SE, each $X$ or $Z$ error in the data qubits will lead to a pair of defects in the bulk (with singular defects at the boundaries). Provided no more than $t$ errors occurred, the defect locations across the $d^2-1$ syndrome measurement results pinpoint the location of the errors, which can then be removed. In the presence of SE imperfections, the syndrome measurements across the patch are carried out $d$ times in succession, and the whole three-dimensional (3D) syndrome volume, comprising $d$ temporal layers of $d^2-1$ (spatial) syndrome measurement results, is analyzed together in the decoding algorithm to deduce the correct (still assuming $\leq t$ faults) data-qubit errors. A logical $X$ or $Z$ error occurs when the actual errors, together with our guessed recovery operations, form a chain that span the length of the surface-code patch, possible---for a distance-preserving decoder---only if there were $>t$ faults in the 3D spacetime volume of syndrome measurements across $d$ layers and across the whole patch area. 

To facilitate the discussion of our information-efficient decoders below, we distinguish between directions in the surface-code patch that are parallel and orthogonal to the logical $X$ and $Z$ operators. We refer to the direction parallel and orthogonal to the logical $X$ operator as the malignant-$X$ and benign-$X$ directions, respectively; see Fig.~\ref{fig:SurfCodePatch}(a). These terms are motivated by the observation that chains of $X$ errors along the malignant-$X$ direction can accumulate into logical $X$ errors and are hence potentially harmful, while those oriented in the orthogonal direction cause Pauli-frame changes (i.e., flips in the local stabilizer eigenvalues), but do not directly contribute to a logical $X$ error. Likewise, we use malignant-$Z$ and benign-$Z$ directions to refer to the parallel and orthogonal directions to the logical $Z$ operator. For the orientation of the patch as shown in Fig.~\ref{fig:SurfCodePatch}, a row is along the benign-$X$ direction, while a column is along the benign-$Z$ direction.

\section{Information-efficient decoders}
Here, we introduce two different decoders capable of (as we will see later) distance-preserving decoding with less information than the standard surface-code decoder described above which uses all available syndromes across the surface-code patch. We refer to the standard decoder as the ``areal decoder", demanding the transmission of a spatial syndrome information volume that scales with the spatial area $d^2$ of the surface-code patch; in contrast, our information-efficient decoders require spatial syndrome information transmission that scales only with the width $d$, not the area, of the code patch. 

Two remarks are in order, before we describe the decoders: First, for tolerance against syndrome-measurement faults, both areal and information-efficient decoders require $d$ temporal layers of spatial syndrome information to be decoded together. Nevertheless, the spatial, not temporal, syndrome volume enters the communication rate needed for decoding to keep up with the generation of syndrome information. The information-efficient decoders hence require a square-root of the communication rate needed for areal decoders. Second, our decoders make use of standard decoding algorithms, but with a subset of the usual syndrome bits as inputs. Below, our examples employ PyMatching with the Sparse-Blossom variant of MWPM.

\subsection{Row-column decoder}
Our first information-efficient decoder is motivated by the following observation: An even number of $X$ errors in the same row (i.e., along the benign-$X$ direction) commutes with the logical $Z$ operator which extends in the same direction. They thus have no effect on that logical operator. In contrast, an odd number of $X$ errors in the same row will flip the value of logical $Z$. Similar statements hold for $Z$ errors in the same column (benign-$Z$ direction) and the logical $X$ operator. 

The goal of the decoder is thus to enforce an even parity for $X(Z)$ errors along each row(column) in the surface-code patch, by detecting those rows(columns) with an odd number of $X(Z)$ errors, corresponding to an odd number of defects in the neighboring ancillary-qubit rows(columns). We thus aggregate the measurement outcomes of all the $Z$ stabilizers in the same row, or all the $X$ stabilizers in the same column, into a single ``detecting region" \cite{mcewen23}. The aggregated row/column outcome (or detector) corresponds to taking the parity of the measurement outcomes of the individual $Z(X)$ stabilizers within a specific row(column) in the current and preceding syndrome extraction rounds---the temporal comparison identifies the defects, and taking the parity thereafter identifies when there is an odd number of defects. The number of such detecting regions is $2(d-1)$ per syndrome measurement round, with $d-1$ of them dedicated to detecting (odd number of) $X$ errors and $d-1$ for $Z$ errors. 
From the aggregated parity values, standard MWPM decoding is used to pair up those parity values, and diagnosed errors can be attributed to data qubits on the boundary. This would finally give an even number of $X$ and $Z$ errors in every row and column, respectively. 

Now, one may wind up with a long chain of $X$ errors in a column (the malignant-$X$ direction), potentially leading to a logical $X$ error (possibly with additional $X$ errors in subsequent rounds). However, the even parity enforced by our decoding ensures that, if such a long chain exists, there must be a second chain of equal length extending along the same direction. The two chains together commute with the logical $Z$ and thus have no effect on the logical values of the patch.

Note that, in choosing to decode in this row-column manner, we have fixed the logical $X$ and $Z$ operators to be along the vertical and horizontal directions, respectively. Other logical representatives, such as those running along diagonal directions, are not guaranteed to be preserved by our decoding. Nevertheless, one fixed choice of logical $X$ and $Z$ suffices for the code to carry the logical qubit.

\subsection{Boundary decoder}
The row-column decoder uses syndrome information gathered from the standard SE circuits of Fig.~\ref{fig:SurfCodePatch}(c). For our second information-efficient decoder, we rely instead on the dynamic SE circuits introduced in Ref.~\cite{mcewen23}. Dynamic SE circuits give rise to novel features in the error correction process compared to traditional circuits. In traditional SE circuits, stabilizers remain static and in place during the syndrome measurement rounds; in dynamic circuits, the stabilizers evolve by expanding and contracting, or even moving around. This evolution allows the flexibility of changing the code generators after each SE round, and Ref.~\cite{mcewen23} made use of this flexibility to reduce the connectivity requirements for implementing SE for the rotated planar surface codes. Recent experimental work has demonstrated the feasibility of this circuit in hardware \cite{eickbusch2025demonstration}. Dynamic circuits have further been applied to lower connectivity demands in bivariate bicycle codes \cite{shaw_lowering_2025}, and to provide resilience against qubit dropouts \cite{debroy2024luci}.

\begin{figure}
\includegraphics[trim=13mm 104mm 80mm 53mm, clip, width=1.05\columnwidth]{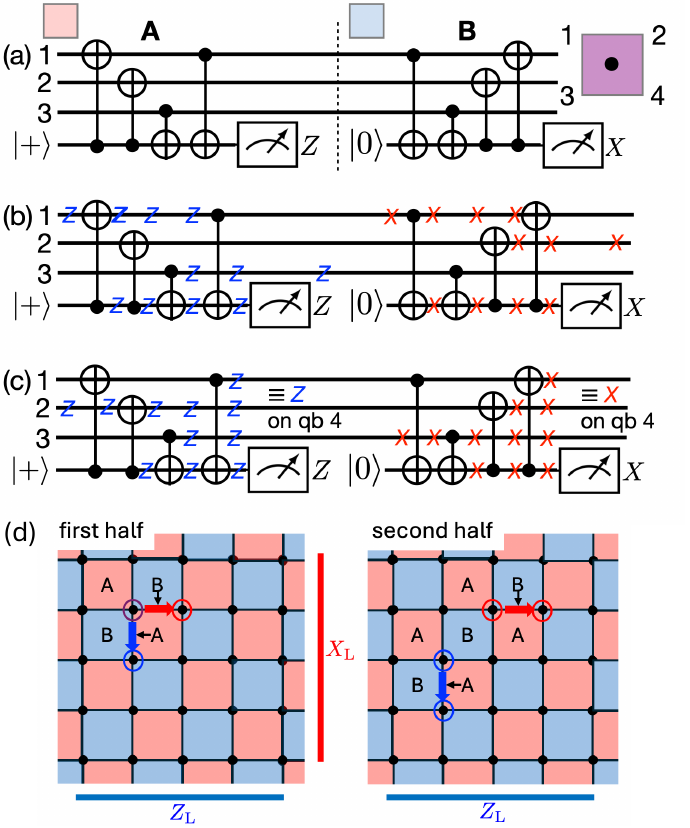}
\caption{\label{fig:3CXCircuit} The 3CX (or hex-grid) syndrome extraction circuit of Ref.~\cite{mcewen23}. (a) The syndrome extraction circuit, comprising two halves, A (left half) and B (right half), applied in alternation to each square in the bulk of the surface-code patch. The purple square on the right gives the data qubit labeling; the ancilla is the black dot in the middle. (b) and (c) show how input errors on the data qubits are moved to different qubits after the A- and B-circuits. The colored $X$s or $Z$s indicate where the error has spread to after the preceding gate operation. Only input errors that are moved in the current SE round are shown. The A- and B-circuits pump errors along the benign directions towards the boundaries of the surface-code patch; see main text for more details. (d) shows an example in the 2D layout for 1 SE round, with a $Y$ error (purple circle) occurring in the bulk just before the start of the round; the two panels show how the $X$ (red circle) and $Z$ (blue circle) errors propagate under the two halves of the SE round under consecutive $A$ and $B$ circuits. The $A/B$ circuit responsible for the movement is indicated by the arrow beside the $A/B$ labels.}
\end{figure}

A curious side effect of the dynamic SE circuits is that errors present before SE can propagate into different errors afterwards. For the SE to not break the requirements of fault tolerance and affect the error-correction performance, dynamic SE circuits should not spread errors uncontrollably. However, they are not constrained to keep the errors in place. In fact, the circuit of Ref.~\cite{mcewen23} has the effect of moving errors along the benign directions towards the boundaries. 

To see this, we look more closely at the dynamic SE circuit proposed in Ref.~\cite{mcewen23}, shown here in Fig.~\ref{fig:3CXCircuit}(a). This is the hex-grid circuit of Ref.~\cite{mcewen23}, referred to elsewhere also as the ``3CX'' circuit \cite{hetenyi24}, since it requires CNOT (or CX) connections between three---not the usual four---data qubits (labeled 1--3 in Fig.~\ref{fig:3CXCircuit}) and the ancilla. For each square in the bulk of the surface-code patch comprising four data qubits at its corners (purple square in top right of Fig.~\ref{fig:3CXCircuit}), the dynamic SE circuit alternates between the A-circuit [left half of Fig.~\ref{fig:3CXCircuit}(a)] and the B-circuit (right half). At initial time, if the square corresponds to a $X(Z)$ stabilizer [red(blue) square in Fig.~\ref{fig:SurfCodePatch}], it begins with the A(B)-circuit; the A- or B-circuits are subsequently applied alternately on those squares. 

Viewed in terms of how the code stabilizers morph as SE proceeds, the A(B)-circuit changes a $X(Z)$-stabilizer [red(blue)] square into a $Z(X)$-stabilizer [blue(red)] square. Each square in the checkerboard pattern of the code patch thus alternates between being an $X$ stabilizer and a $Z$ stabilizer, switching every half cycle of the full SE round (A+B or B+A). The A(B)-circuit is applied whenever the square is an $X(Z)$ stabilizer. Every $X$ or $Z$ stabilizer in the code patch is measured in each round, either in A or B. This structure becomes clearer when analyzed via detecting regions. A detector in a circuit is a set of measurement outcomes with deterministic parity in the absence of errors. The detecting region associated with that detector is the set of circuit locations such that Pauli errors anywhere in the set would flip that parity value. At the beginning of an SE round (A or B), there are $2(d^2-1)$ detecting regions in all, with $d^2-1$ expanding and $d^2-1$ contracting. The expanding detecting regions originate from ancilla resets and evolve to cover the code stabilizers by the end of the SE round. Conversely, the contracting detecting regions are those that began expanding in the previous round and terminate in ancilla measurements in the current round. At the start of a round, the $d^2-1$ contracting regions collectively cover the $d^2-1$ stabilizer generators of the code. By the end of the round, the measurements of the ancillary qubits within those regions yield the eigenvalues of the stabilizers. See Ref.~\cite{mcewen23} for more details.

An alternate way, more pertinent for our discussion here, to understand this 3CX circuit is to look at how the errors are modified by the dynamic SE circuit. Figures \ref{fig:3CXCircuit}(b) and (c) show the four distinct situations that can arise, with a $Z$ input error on data qubit 1 or 2, or an $X$ input error on data qubit 1 or 3. From Fig.~\ref{fig:3CXCircuit}(b), we see that a $Z$ input error on data qubit 1 turns into a $Z$ error on qubit 3 at the completion of the A-circuit; for a $Z$ input error on qubit 2 [see Fig.~\ref{fig:3CXCircuit}(c)], the A-circuit turns it into $ZZZ$ on qubits 1--3, equivalent---modulo the code stabilizer at that spacetime location (a $Z$-stabilizer)---to a single $Z$ error on qubit 4; similar statements hold for $X$ input errors. Since every data qubit is qubit 1--4 for four different squares, the four situations in Figs.~\ref{fig:3CXCircuit}(b) and (c) cover all possibilities.

From Figs.~\ref{fig:3CXCircuit}(b) and (c), we see that the dynamic SE circuits move the errors. Specifically, they ``pump'' $X$ errors rightwards---along the benign-$X$ direction---by one unit (e.g., qubit 1 to 2) every SE round; at the same time, $Z$ errors are pumped downwards---along the benign-$Z$ direction---by one unit (e.g., qubit 1 to 3). Eventually, an error originating anywhere in surface-code patch reaches the right or bottom boundaries of the patch. At the boundary, the stabilizers and SE circuits for the boundary data qubits have to be chosen carefully, as highlighted in Ref.~\cite{mcewen23}; we leave the reader to consult the original reference for the full details. Of relevance for us here is the fact that the specific choice in Ref.~\cite{mcewen23} causes errors that arrive at the right and bottom boundaries to remain there, unchanged by subsequent SE rounds, except when an identical error arrives at the same location and neutralizes an existing error there. These error movements are visualized in a video available online \cite{my_2025_17851272}.

\begin{figure}
    \centering
    \includegraphics[trim=10mm 120mm 120mm 8mm, clip, width=0.5\columnwidth]{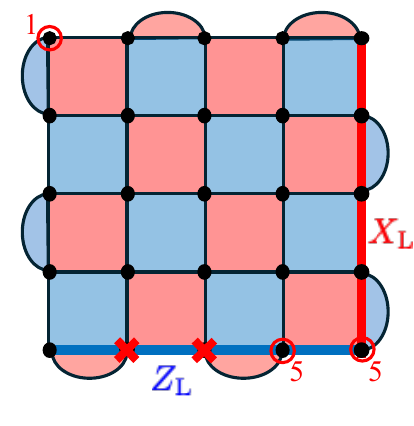}
    \caption{Example of a logical error from $>t$ faults in the absence of benign-direction information. A $d=5$ example is given here, with actual $Z$ errors occurring on the top-left qubit at the beginning of SE round 1 (marked with a red circle labeled `1') and two more on the bottom boundary at the beginning of round 5 (red circles labeled `5'). With the 3CX SE circuit, the error on the top-left qubit arrives at the bottom boundary after 4 SE rounds. That, and the two errors occurring in round 5, trigger two boundary $X$-detectors in round 5. Based on these boundary detection events, the boundary decoder diagnoses the two qubits marked with red crosses to have errors. This recovery, with the actual errors, lead to a logical error. If the areal decoder is used instead, the benign-direction information will properly identify the error originating in the top-left to have occurred in round 1, with no confusion with the errors occurring much later in the bottom right.}
    \label{fig:MalignantError}
\end{figure}

This pumping of errors along the benign directions to the boundaries provides the basis for our information-efficient boundary decoder: By collecting only the syndrome data along the bottom and right boundaries, we have the necessary data to remove the errors. $Z$ errors are pumped in the benign-$Z$ direction, so the spatial separation along the malignant-$Z$ direction between every pair of $Z$ errors is maintained. Similar statements hold for the $X$ errors.
Information about temporal separation and spatial separation in the benign direction, however, become confounded with one another. This ambiguity is harmless when the number of errors is within the code distance---it cannot lead to a recovery that results in a logical error; one simply does not have enough errors to cause a logical error, whether or not we know that benign-direction separation. Thus, our decoder will function correctly with just syndrome information from the bottom and right boundaries, gathered over multiple SE rounds. 

The intuition from the earlier row-column decoder is also useful here: By monitoring the boundary information, we know if an even or odd number of errors have arrived there, which is all that is needed to decode the logical values. Note, however, that when one has errors beyond the code distance, the benign-direction information can help distinguish between different ways to match up the detected defects. Our boundary decoder, lacking that information, hence cannot be expected to do as well as the standard areal decoder in such situations. An example of a logical error that can arise due to the loss of benign-direction information is given in Fig.~\ref{fig:MalignantError}.

\section{Decoder performance}

Our inspiration and rationale for the information-efficient decoders so far are largely based on error-free SE. To verify decoding accuracy even for noisy SE, as well as to demonstrate the performance of our decoders, we study the different decoders numerically through memory experiments in the presence of circuit-level noise. 

\begin{figure*}
    \centering
    \includegraphics[width=\textwidth]{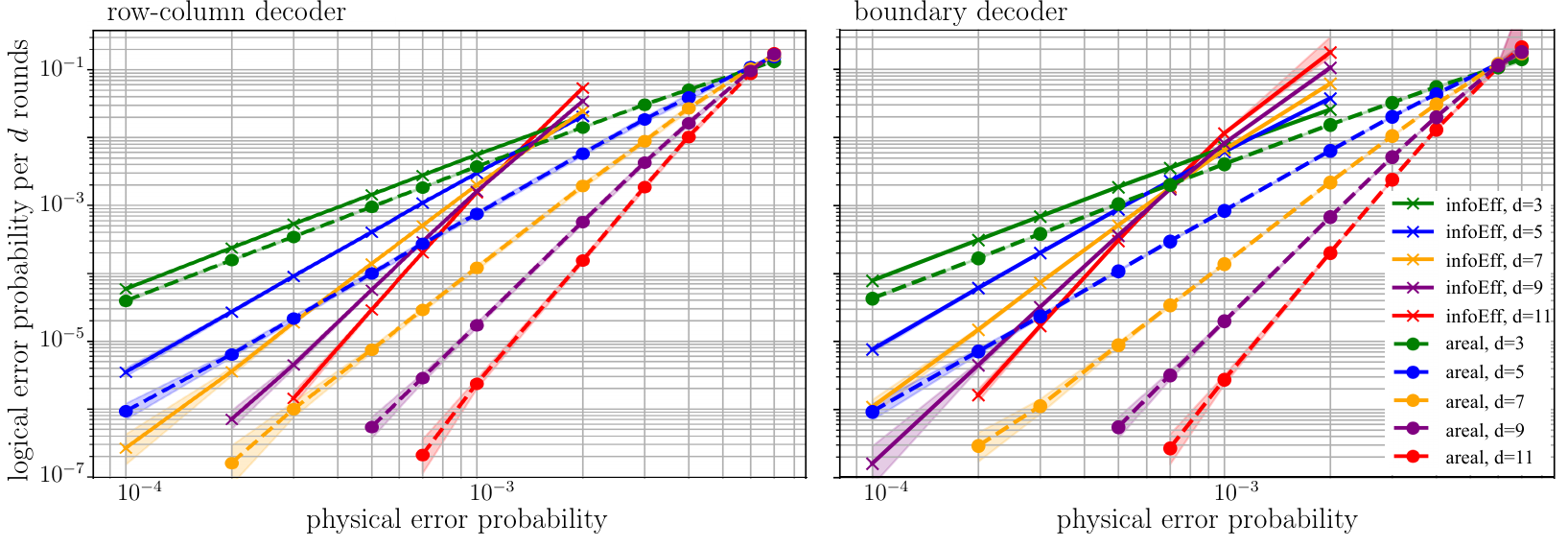}
    \caption{Performance of the information-efficient decoders in numerical memory experiments with circuit-level noise, for surface code of distances $d=3,5,\ldots,11$. On the left, we have the row-column decoder (solid lines and cross markers) compared with the standard areal decoder (dashed lines with dot markers), for the standard syndrome extraction circuit; on the right, we have the boundary decoder compared with the full areal decoder, now for the 3CX syndrome extraction circuit.}
    \label{fig:Results}
\end{figure*}

Now, in actual use in a quantum computation, we expect our information-efficient decoders to handle the decoding in the middle of the computation, after the state initialization and before the final measurement of the surface-code patch. For the state initialization, involving typically the first $d$ rounds of SE in an experiment, full areal decoding is needed to stabilize the code space into one eigenspace of the code stabilizers; for the final measurement of the surface-code patch, errors scattered across the patch that occur during these final SE rounds may not all be caught by the information-efficient decoders, so again, we apply the areal decoder there. These initial and final uses of the areal decoder do not scale with the length of the computation; during the computation, we employ the information-efficient decoders to reduce the communication throughput.

Our numerical experiment here is thus designed to focus on the efficacy of the information-efficient decoders in the middle of the computation. We implement ideal encoding by initializing all physical qubits in the the $\ket{0}$ or $\ket{+}$ states, followed by noise-free SE which projects the patch into the code space. We then go through 20$d$ rounds of SE, with all physical operations subject to depolarizing noise of strength $p$ (i.e., Pauli errors each occur probability $p/3$ for single-qubit operations and $p/15$ for two-qubit ones). Finally, we conclude with a single round of ideal SE, followed by $X$ or $Z$ measurements, depending on whether we started with $\ket{0}$ or $\ket{+}$, and thus compute the logical error probability for varying $p$ values. We perform the simulations with two input states $\ket{0}_\mathrm{L}$ and the other with $\ket{+}_\mathrm{L}$, giving the logical $Z$ and $X$ error probabilities $p_{\mathrm{L}_Z}$ and $p_{\mathrm{L}_X}$. The total logical error probability is the combination, $p_\mathrm{L}=p_{\mathrm{L}_Z}+p_{\mathrm{L}_X}-p_{\mathrm{L}_Z}p_{\mathrm{L}_X}$, assuming the $X$ and $Z$ errors are independent at the logical level.

The memory experiments were simulated using Stim \cite{gidney21}, with PyMatching \cite{higgott22} (with Sparse Blossom \cite{higgott25}) as the decoding algorithm for the collected syndrome information. For the row-column decoder, only the aggregated row/column parity values were sent to the decoder; detectors were formed by comparing the current round's parity values with those from the same row/column in the previous SE round. For the boundary decoder, only the syndrome bits on the right and bottom boundaries were sent to the decoder; detectors were formed by comparing the current syndrome outcome at a boundary location with the syndrome value at the same location in the previous round. 

Now, the noise-free SE in the initialization step puts the code space into a stabilizer eigenspace with random assignment of $+1$ and $-1$ stabilizer-generator eigenvalues (measurement outcomes) across the surface-code patch. Because the boundary decoder sees only the eigenvalues at the right and bottom boundaries in each round, and is unaware of the eigenvalues elsewhere in the patch, a $-1$ eigenvalue elsewhere will appear like an error that gets eventually pumped to the right/bottom boundary in subsequent SE rounds. To avoid such mis-identification while not having to define additional areal detectors in Stim, the initialization stage for the boundary decoder case involves $d$ rounds of noise-free SE, to propagate the $-1$ eigenvalues to the boundaries, before carrying out the $20d$ rounds of SE in the memory experiment. No such extra noise-free SE rounds are needed for the row/column decoder.

Figure \ref{fig:Results} shows the logical versus physical error probabilities for our information efficient decoders (cross markers); we give also the logical error rates for the standard areal decoder (dot markers). From the plots, the identical gradients of the same-$d$ lines for the all three decoders tell us that our information-efficient decoders preserve the code distances as promised by the surface code. However, the significantly higher $y$-intercept values, and hence higher logical error probabilities, for our information-efficient decoders signal the inability to cope with as many of the $>\!\!\lfloor (d-1)/2\rfloor$-fault situations as the areal decoder. This is expected from our discussion above, due to the loss of information in the benign directions for information efficiency. We observe a corresponding order-of-magnitude worsening of the threshold values for our information decoders, compared to the standard areal decoder.

To further confirm that our information-decoders nevertheless preserve the code distance, we made use of Stim's $\texttt{shortest\_graphlike\_error()}$ function to check that the circuit used for the distance-$d$ code indeed has circuit distance $d$, i.e., the minimum number of faults anywhere in the circuit needed to flip the logical observable value without triggering any detectors is $d$. Then, for the $d=3$ and $5$ codes, with just $4d$ rounds of SE (shortening the earlier $20d$ rounds) we exhaustively verify that our decoders correctly removes all 1-fault and 2-fault situations, respectively, anywhere in the circuit (provided the physical error probability is small enough for MWPM decoding to work).

\section{Conclusion}

We have described two decoders that use only a reduced set of the spatial syndrome information. 
Despite the higher logical error probability compared to the areal decoder, our information-efficient decoders give a significant reduction in the amount of information that has to be transmitted out to the classical decoding algorithm, from $O(d^2)$ per SE round for the areal decoder to $O(d)$ for the information-efficient decoders, alleviating the demands on the classical communication rate needed to extract the syndrome information to the decoder on the CPU. Furthermore, the size of the decoding graph used in PyMatching is also correspondingly reduced from $O(d^2)$ to $O(d)$, leading to a significant speedup in the decoding runtime, from $O(d^6\log(d))$ to $O(d^3\log(d))$ \cite{higgott22}. 

Even though we use only the syndrome bits from two of the patch boundaries in the boundary decoder, the SE operations still have to be done across the entire surface, with those operations now playing the role of pumping the errors to the boundaries. The current decoder, with the 3CX SE circuit, gives a notably lower logical error probability; nevertheless, this raises the intriguing question of whether there are other error-pumping circuits that an move the errors in different ways that lead to more self-cancellations and hence improve the logical error probability.

\acknowledgements
HK Ng acknowledges partial support from the National Research Foundation, Singapore through the National Quantum Office, hosted in A*STAR, under its Centre for Quantum Technologies Funding Initiative (S24Q2d0009).

\bibliographystyle{quantum}
\bibliography{InfoEffDecoding.bib}

@article{zhang_classical_2023,
	title = {A {Classical} {Architecture} for {Digital} {Quantum} {Computers}},
	volume = {5},
	url = {https://dl.acm.org/doi/10.1145/3626199},
	doi = {10.1145/3626199},
	number = {1},
	urldate = {2025-06-09},
	journal = {ACM Transactions on Quantum Computing},
	author = {Zhang, Fang and Zhu, Xing and Chao, Rui and Huang, Cupjin and Kong, Linghang and Chen, Guoyang and Ding, Dawei and Feng, Haishan and Gao, Yihuai and Ni, Xiaotong and Qiu, Liwei and Wei, Zhe and Yang, Yueming and Zhao, Yang and Shi, Yaoyun and Zhang, Weifeng and Zhou, Peng and Chen, Jianxin},
	month = dec,
	year = {2023},
	pages = {3:1--3:24},
}

@inproceedings{das_afs_2022,
	address = {Seoul, Korea, Republic of},
	title = {{AFS}: {Accurate}, {Fast}, and {Scalable} {Error}-{Decoding} for {Fault}-{Tolerant} {Quantum} {Computers}},
	copyright = {https://doi.org/10.15223/policy-029},
	isbn = {978-1-6654-2027-3},
	shorttitle = {{AFS}},
	url = {https://ieeexplore.ieee.org/document/9773217/},
	doi = {10.1109/HPCA53966.2022.00027},
	language = {en},
	urldate = {2025-06-09},
	booktitle = {2022 {IEEE} {International} {Symposium} on {High}-{Performance} {Computer} {Architecture} ({HPCA})},
	publisher = {IEEE},
	author = {Das, Poulami and Pattison, Christopher A. and Manne, Srilatha and Carmean, Douglas M. and Svore, Krysta M. and Qureshi, Moinuddin and Delfosse, Nicolas},
	month = apr,
	year = {2022},
	pages = {259--273},
}

@inproceedings{camps_evaluation_2024,
	title = {Evaluation of the {Classical} {Hardware} {Requirements} for {Large}-{Scale} {Quantum} {Computations}},
	url = {https://ieeexplore.ieee.org/document/10528937},
	doi = {10.23919/ISC.2024.10528937},
	urldate = {2025-06-09},
	booktitle = {{ISC} {High} {Performance} 2024 {Research} {Paper} {Proceedings} (39th {International} {Conference})},
	author = {Camps, Daan and Rrapaj, Ermal and Klymko, Katherine and Austin, Brian and Wright, Nicholas J.},
	month = may,
	year = {2024},
	pages = {1--12},
}

@article{mcewen23,
	title = {Relaxing {Hardware} {Requirements} for {Surface} {Code} {Circuits} using {Time}-dynamics},
	volume = {7},
	url = {https://quantum-journal.org/papers/q-2023-11-07-1172/},
	doi = {10.22331/q-2023-11-07-1172},
	language = {en-GB},
	urldate = {2025-06-12},
	journal = {Quantum},
	author = {McEwen, Matt and Bacon, Dave and Gidney, Craig},
	month = nov,
	year = {2023},
	note = {Publisher: Verein zur Förderung des Open Access Publizierens in den Quantenwissenschaften},
	pages = {1172},
}

@article{shaw_lowering_2025,
	title = {Lowering {Connectivity} {Requirements} for {Bivariate} {Bicycle} {Codes} {Using} {Morphing} {Circuits}},
	volume = {134},
	issn = {0031-9007, 1079-7114},
	url = {https://link.aps.org/doi/10.1103/PhysRevLett.134.090602},
	doi = {10.1103/PhysRevLett.134.090602},
	language = {en},
	number = {9},
	urldate = {2025-06-12},
	journal = {Physical Review Letters},
	author = {Shaw, Mackenzie H. and Terhal, Barbara M.},
	month = mar,
	year = {2025},
	pages = {090602},
}

@article{higgott22,
author = {Higgott, Oscar},
title = {PyMatching: A Python Package for Decoding Quantum Codes with Minimum-Weight Perfect Matching},
year = {2022},
issue_date = {September 2022},
publisher = {Association for Computing Machinery},
address = {New York, NY, USA},
volume = {3},
number = {3},
url = {https://doi.org/10.1145/3505637},
doi = {10.1145/3505637},
journal = {ACM Transactions on Quantum Computing~},
month = {jun},
articleno = {16},
numpages = {16}
}

@article{kitaev97,
author = {Kitaev, A. Yu.},
title = {Quantum computations: algorithms and error correction},
year = {1997},
volume = {52},
number = {6},
url = {https://doi.org/10.1070/RM1997v052n06ABEH002155},
doi = {10.1070/RM1997v052n06ABEH002155},
journal = {Russian Math. Surveys},
pages = {1191--1249},
}

@article{kitaev03,
title = {Fault-tolerant quantum computation by anyons},
journal = {Annals of Physics},
volume = {303},
number = {1},
pages = {2-30},
year = {2003},
issn = {0003-4916},
doi = {https://doi.org/10.1016/S0003-4916(02)00018-0},
url = {https://www.sciencedirect.com/science/article/pii/S0003491602000180},
author = {A.Yu. Kitaev},
note = {arXiv:quant-ph/9707021},
}

@article{dennis02,
	title = {Topological quantum memory},
	volume = {43},
	issn = {0022-2488},
	url = {https://doi.org/10.1063/1.1499754},
	doi = {10.1063/1.1499754},
	number = {9},
	urldate = {2025-08-01},
	journal = {Journal of Mathematical Physics},
	author = {Dennis, Eric and Kitaev, Alexei and Landahl, Andrew and Preskill, John},
	month = sep,
	year = {2002},
	pages = {4452--4505},
}

@misc{bravyi98,
      title={Quantum codes on a lattice with boundary}, 
      author={S. B. Bravyi and A. Yu. Kitaev},
      year={1998},
      eprint={quant-ph/9811052},
      archivePrefix={arXiv},
      primaryClass={quant-ph},
      url={https://arxiv.org/abs/quant-ph/9811052}, 
}

@article{fowler12,
  title = {Surface codes: Towards practical large-scale quantum computation},
  author = {Fowler, Austin G. and Mariantoni, Matteo and Martinis, John M. and Cleland, Andrew N.},
  journal = {Phys. Rev. A},
  volume = {86},
  issue = {3},
  pages = {032324},
  numpages = {48},
  year = {2012},
  month = {Sep},
  publisher = {American Physical Society},
  doi = {10.1103/PhysRevA.86.032324},
  url = {https://link.aps.org/doi/10.1103/PhysRevA.86.032324}
}

@article{gidney21,
  doi = {10.22331/q-2021-07-06-497},
  url = {https://doi.org/10.22331/q-2021-07-06-497},
  title = {Stim: a fast stabilizer circuit simulator},
  author = {Gidney, Craig},
  journal = {{Quantum}},
  issn = {2521-327X},
  publisher = {{Verein zur F{\"{o}}rderung des Open Access Publizierens in den Quantenwissenschaften}},
  volume = {5},
  pages = {497},
  month = jul,
  year = {2021}
}

@article{higgott25,
  doi = {10.22331/q-2025-01-20-1600},
  url = {https://doi.org/10.22331/q-2025-01-20-1600},
  title = {Sparse {B}lossom: correcting a million errors per core second with minimum-weight matching},
  author = {Higgott, Oscar and Gidney, Craig},
  journal = {{Quantum}},
  issn = {2521-327X},
  publisher = {{Verein zur F{\"{o}}rderung des Open Access Publizierens in den Quantenwissenschaften}},
  volume = {9},
  pages = {1600},
  month = jan,
  year = {2025}
}

@article{Hetenyi24,
  title = {Creating Entangled Logical Qubits in the Heavy-Hex Lattice with Topological Codes},
  author = {Het\'enyi, Bence and Wootton, James R.},
  journal = {PRX Quantum},
  volume = {5},
  issue = {4},
  pages = {040334},
  numpages = {14},
  year = {2024},
  month = {Dec},
  publisher = {American Physical Society},
  doi = {10.1103/PRXQuantum.5.040334},
  url = {https://link.aps.org/doi/10.1103/PRXQuantum.5.040334}
}

@article{google2023suppressing,
  title={Suppressing quantum errors by scaling a surface code logical qubit},
  author={Google Quantum AI},
  journal={Nature},
  volume={614},
  number={7949},
  pages={676--681},
  year={2023},
  publisher={Nature Publishing Group UK London},
    url={https://doi.org/10.1038/s41586-022-05434-1}
}

@article{google2025quantum,
  title={Quantum error correction below the surface code threshold},
author={Google Quantum AI and Collaborators},
  journal={Nature},
  volume={638},
  number={8052},
  pages={920--926},
  year={2025},
  publisher={Nature Publishing Group UK London},
url={https://doi.org/10.1038/s41586-024-08449-y}
}

@article{eickbusch2025demonstration,
  title={Demonstration of dynamic surface codes},
  author={Eickbusch, Alec and McEwen, Matt and Sivak, Volodymyr and Bourassa, Alexandre and Atalaya, Juan and Claes, Jahan and Kafri, Dvir and Gidney, Craig and Warren, Christopher W and Gross, Jonathan and others},
  journal={Nature Physics},
  pages={1--8},
  year={2025},
  publisher={Nature Publishing Group UK London},
url={https://doi.org/10.1038/s41567-025-03070-w}
}

@article{bluvstein2024logical,
  title={Logical quantum processor based on reconfigurable atom arrays},
  author={Bluvstein, Dolev and Evered, Simon J and Geim, Alexandra A and Li, Sophie H and Zhou, Hengyun and Manovitz, Tom and Ebadi, Sepehr and Cain, Madelyn and Kalinowski, Marcin and Hangleiter, Dominik and others},
  journal={Nature},
  volume={626},
  number={7997},
  pages={58--65},
  year={2024},
  publisher={Nature Publishing Group UK London},
url={https://doi.org/10.1038/s41586-023-06927-3}
}

@article{bluvstein2025architectural,
  title={Architectural mechanisms of a universal fault-tolerant quantum computer},
  author={Bluvstein, Dolev and Geim, Alexandra A and Li, Sophie H and Evered, Simon J and Ataides, J and Baranes, Gefen and Gu, Andi and Manovitz, Tom and Xu, Muqing and Kalinowski, Marcin and others},
  journal={arXiv preprint arXiv:2506.20661},
  year={2025},
url={https://doi.org/10.48550/arXiv.2506.20661}
}

@article{bravyi2024high,
  title={High-threshold and low-overhead fault-tolerant quantum memory},
  author={Bravyi, Sergey and Cross, Andrew W and Gambetta, Jay M and Maslov, Dmitri and Rall, Patrick and Yoder, Theodore J},
  journal={Nature},
  volume={627},
  number={8005},
  pages={778--782},
  year={2024},
  publisher={Nature Publishing Group UK London},
url={https://doi.org/10.1038/s41586-024-07107-7}
}

@article{xu2024constant,
  title={Constant-overhead fault-tolerant quantum computation with reconfigurable atom arrays},
  author={Xu, Qian and Bonilla Ataides, J Pablo and Pattison, Christopher A and Raveendran, Nithin and Bluvstein, Dolev and Wurtz, Jonathan and Vasi{\'c}, Bane and Lukin, Mikhail D and Jiang, Liang and Zhou, Hengyun},
  journal={Nature Physics},
  volume={20},
  number={7},
  pages={1084--1090},
  year={2024},
  publisher={Nature Publishing Group UK London},
url={https://doi.org/10.1038/s41567-024-02479-z}
}

@article{pattison2025hierarchical,
  title={Hierarchical memories: Simulating quantum LDPC codes with local gates},
  author={Pattison, Christopher A and Krishna, Anirudh and Preskill, John},
  journal={Quantum},
  volume={9},
  pages={1728},
  year={2025},
  publisher={Verein zur F{\"o}rderung des Open Access Publizierens in den Quantenwissenschaften},
url={https://doi.org/10.22331/q-2025-05-05-1728}
}

@article{terhal2015quantum,
  title={Quantum error correction for quantum memories},
  author={Terhal, Barbara M},
  journal={Reviews of Modern Physics},
  volume={87},
  number={2},
  pages={307--346},
  year={2015},
  publisher={APS},
url={https://doi.org/10.1103/RevModPhys.87.307}
}

@article{debroy2024luci,
  title={LUCI in the Surface Code with Dropouts},
  author={Debroy, Dripto M and McEwen, Matt and Gidney, Craig and Shutty, Noah and Zalcman, Adam},
  journal={arXiv preprint arXiv:2410.14891},
  year={2024},
url={
https://doi.org/10.48550/arXiv.2410.14891}
}

@article{berthusen2025toward,
  title={Toward a 2D local implementation of quantum low-density parity-check codes},
  author={Berthusen, Noah and Devulapalli, Dhruv and Schoute, Eddie and Childs, Andrew M and Gullans, Michael J and Gorshkov, Alexey V and Gottesman, Daniel},
  journal={PRX Quantum},
  volume={6},
  number={1},
  pages={010306},
  year={2025},
  publisher={APS},
url={
https://doi.org/10.1103/PRXQuantum.6.010306}
}

@article{skoric2023parallel,
  title={Parallel window decoding enables scalable fault tolerant quantum computation},
  author={Skoric, Luka and Browne, Dan E and Barnes, Kenton M and Gillespie, Neil I and Campbell, Earl T},
  journal={Nature Communications},
  volume={14},
  number={1},
  pages={7040},
  year={2023},
  publisher={Nature Publishing Group UK London},
url={https://doi.org/10.1038/s41467-023-42482-1}
}

@article{tan2023scalable,
  title={Scalable surface-code decoders with parallelization in time},
  author={Tan, Xinyu and Zhang, Fang and Chao, Rui and Shi, Yaoyun and Chen, Jianxin},
  journal={PRX Quantum},
  volume={4},
  number={4},
  pages={040344},
  year={2023},
  publisher={APS},
url={https://doi.org/10.1103/PRXQuantum.4.040344}
}

@article{breuckmann2021,
  title = {Quantum Low-Density Parity-Check Codes},
  author = {Breuckmann, Nikolas P. and Eberhardt, Jens Niklas},
  journal = {PRX Quantum},
  volume = {2},
  issue = {4},
  pages = {040101},
  numpages = {19},
  year = {2021},
  month = {Oct},
  publisher = {American Physical Society},
  doi = {10.1103/PRXQuantum.2.040101},
  url = {https://link.aps.org/doi/10.1103/PRXQuantum.2.040101}
}

@article{edmonds1965, 
title={Paths, Trees, and Flowers}, 
volume={17}, DOI={10.4153/CJM-1965-045-4}, 
journal={Canadian Journal of Mathematics}, 
author={Edmonds, Jack}, 
year={1965}, pages={449–467}}

@article{delfosse2021,
  doi = {10.22331/q-2021-12-02-595},
  url = {https://doi.org/10.22331/q-2021-12-02-595},
  title = {Almost-linear time decoding algorithm for topological codes},
  author = {Delfosse, Nicolas and Nickerson, Naomi H.},
  journal = {{Quantum}},
  issn = {2521-327X},
  publisher = {{Verein zur F{\"{o}}rderung des Open Access Publizierens in den Quantenwissenschaften}},
  volume = {5},
  pages = {595},
  month = dec,
  year = {2021}
}

@article{bravyi2014,
  title = {Efficient algorithms for maximum likelihood decoding in the surface code},
  author = {Bravyi, Sergey and Suchara, Martin and Vargo, Alexander},
  journal = {Phys. Rev. A},
  volume = {90},
  issue = {3},
  pages = {032326},
  numpages = {15},
  year = {2014},
  month = {Sep},
  publisher = {American Physical Society},
  doi = {10.1103/PhysRevA.90.032326},
  url = {https://link.aps.org/doi/10.1103/PhysRevA.90.032326}
}

@article{torlai2017,
  title = {Neural Decoder for Topological Codes},
  author = {Torlai, Giacomo and Melko, Roger G.},
  journal = {Phys. Rev. Lett.},
  volume = {119},
  issue = {3},
  pages = {030501},
  numpages = {5},
  year = {2017},
  month = {Jul},
  publisher = {American Physical Society},
  doi = {10.1103/PhysRevLett.119.030501},
  url = {https://link.aps.org/doi/10.1103/PhysRevLett.119.030501}
}

@article{krastanov2017,
	author = {Krastanov, Stefan and Jiang, Liang},
	date = {2017/09/08},
	doi = {10.1038/s41598-017-11266-1},
	id = {Krastanov2017},
	isbn = {2045-2322},
	journal = {Scientific Reports},
	number = {1},
	pages = {11003},
	title = {Deep Neural Network Probabilistic Decoder for Stabilizer Codes},
	url = {https://doi.org/10.1038/s41598-017-11266-1},
	volume = {7},
	year = {2017}}

@article{iOlius2024,
  doi = {10.22331/q-2024-10-10-1498},
  url = {https://doi.org/10.22331/q-2024-10-10-1498},
  title = {Decoding algorithms for surface codes},
  author = {deMarti iOlius, Antonio and Fuentes, Patricio and Or{\'{u}}s, Rom{\'{a}}n and Crespo, Pedro M. and Etxezarreta Martinez, Josu},
  journal = {{Quantum}},
  issn = {2521-327X},
  publisher = {{Verein zur F{\"{o}}rderung des Open Access Publizierens in den Quantenwissenschaften}},
  volume = {8},
  pages = {1498},
  month = oct,
  year = {2024}
}

@article{bravyi2005,
  title = {Universal quantum computation with ideal Clifford gates and noisy ancillas},
  author = {Bravyi, Sergey and Kitaev, Alexei},
  journal = {Phys. Rev. A},
  volume = {71},
  issue = {2},
  pages = {022316},
  numpages = {14},
  year = {2005},
  month = {Feb},
  publisher = {American Physical Society},
  doi = {10.1103/PhysRevA.71.022316},
  url = {https://link.aps.org/doi/10.1103/PhysRevA.71.022316}
}

@misc{my_2025_17851272,
author = {My, DH Long and Chiew, Shao-Hen and Chai, Jing Hao and Ng, Hui Khoon},
title = {Data for "Information-efficient decoding of surface codes"},
month = dec,
year  = 2025,
publisher = {Zenodo},
doi = {10.5281/zenodo.17851272},
url = {https://doi.org/10.5281/zenodo.17851272},
howpublished = {\url{https://doi.org/10.5281/zenodo.17851272}}
}

\end{document}